\pdfoutput=1

\documentclass[sigconf]{acmart}

\usepackage{booktabs} 
\usepackage{amsmath} 
\usepackage{subfigure} 
\usepackage{graphicx}
\renewcommand\footnotetextcopyrightpermission[1]{} 

\setcopyright{rightsretained}

\begin{document}
\title[Route Recommender with Status Prediction and Safe Routing]{SAFEBIKE: A Bike-sharing Route Recommender with Availability Prediction and Safe Routing}


\author{Weisheng Zhong, Fanglan Chen, Kaiqun Fu, Chang-Tien Lu}
\affiliation{
  \institution{Virginia Tech}
  \streetaddress{7054 Haycock Road}
  \city{Falls Church}
  \state{Virginia}
  \postcode{22043}
}
\affiliation{
	\institution{\{zwscn123, fanglanc, fukaiqun, ctlu\}@vt.edu}
}

\begin{abstract}
This paper presents SAFEBIKE, a novel route recommendation system for bike-sharing service that utilizes station information to infer the number of available bikes in dock and recommend bike routes according to multiple factors such as distance and safety level. The system consists of a station level availability predictor that predicts bikes and docks amount at each station, and an efficient route recommendation service that considers safety and bike/dock availability factors. It targets users who are concerned about route safeness and station availability. We demonstrate the system by utilizing Citi Bike station availability and New York City crime data of Manhattan to show the effectiveness of our approach. Integrated with real-time station availability and historical crime data resources, our proposed system can effectively recommend an optimal bike route and improve the travel experience of bike users.
\end{abstract}

\maketitle

\section{Introduction}
With an increasing demand of public transportation, nowadays many cities worldwide has developed their own bike-sharing systems (BSSs), including France (Paris and Marseilles), China (Beijing and Hangzhou), United States (Chicago, New York and Washington D. C.)\cite{wiki} . Bike-sharing system is a public transportation service in which bikes are made available to individuals for shared use on a short term basis. Bike-sharing system allows people to check-out a bike at a nearby station and return the bike to a station close to their destination. As stations are located in different areas in the city, the usage of bikes among stations is imbalanced\cite{borgnat2011shared}. However, there are two major challenges for the existing bike-sharing systems: 1) At different times of a day, some stations may lack available bikes for individuals to check out, while some may not have empty docks for individuals to drop off bikes. This issue results in individuals take extra cost on traveling to alternative stations or give up using bike-sharing systems, which eventually impair the user experience. 2) On the other hand, as biking is an outdoor activity, good community security is an important factor that affects individuals to choose bike-sharing as their mode of commute\cite{fishman2016global}. To determine the optimal tread-off between travel distance and safety is one of the key requirement for the bike-sharing systems. 

Inspired and motivated by these problems, we developed the SAFEBIKE system, a user-friendly web application that provides real-time bike station availability prediction and an optimal routes recommendation service. It mines station availability information from a bike-sharing system and implements a availability prediction algorithm to forecast bikes and docks availability for each station. The proposed SAFEBIKE system provides the prediction of the number of available bikes for certain stations in the short future. Furthermore, a route recommendation service is developed that combines the predicted  results and crime data to provide optimal routes between two locations within a city. We utilize Citi Bike system's station availability data and crime data for the Manhattan area. 


The major contributions of SAFEBIKE can be summarized as follows:
\begin{itemize}
    \item \textbf{Real-time prediction for station availability: } The system applies a real-time station availability prediction algorithm to forecast the number of bikes and docks at each station for future time periods. 
    \item \textbf{Road Safety Evaluation: } The system utilizes crime data to measure the safeness of roads within a city. A spatial database is established to maintain the large amount of city crime records.
    \item \textbf{Route recommendation service:}  SAFEBIKE provides an optimal route for an individual user based on his or her preferences. Implementation of this feature is based on the predicted availability results and crime data.
    \item \textbf{Interactive interface:} A user-friendly web application interface is developed to incorporate all the above functions. It utilizes several state-of-art web technologies to bring convenience and efficiency to bike-sharing users.
\end{itemize}
\begin{figure}[!htb]
\centering
\includegraphics[width=1.0\linewidth]{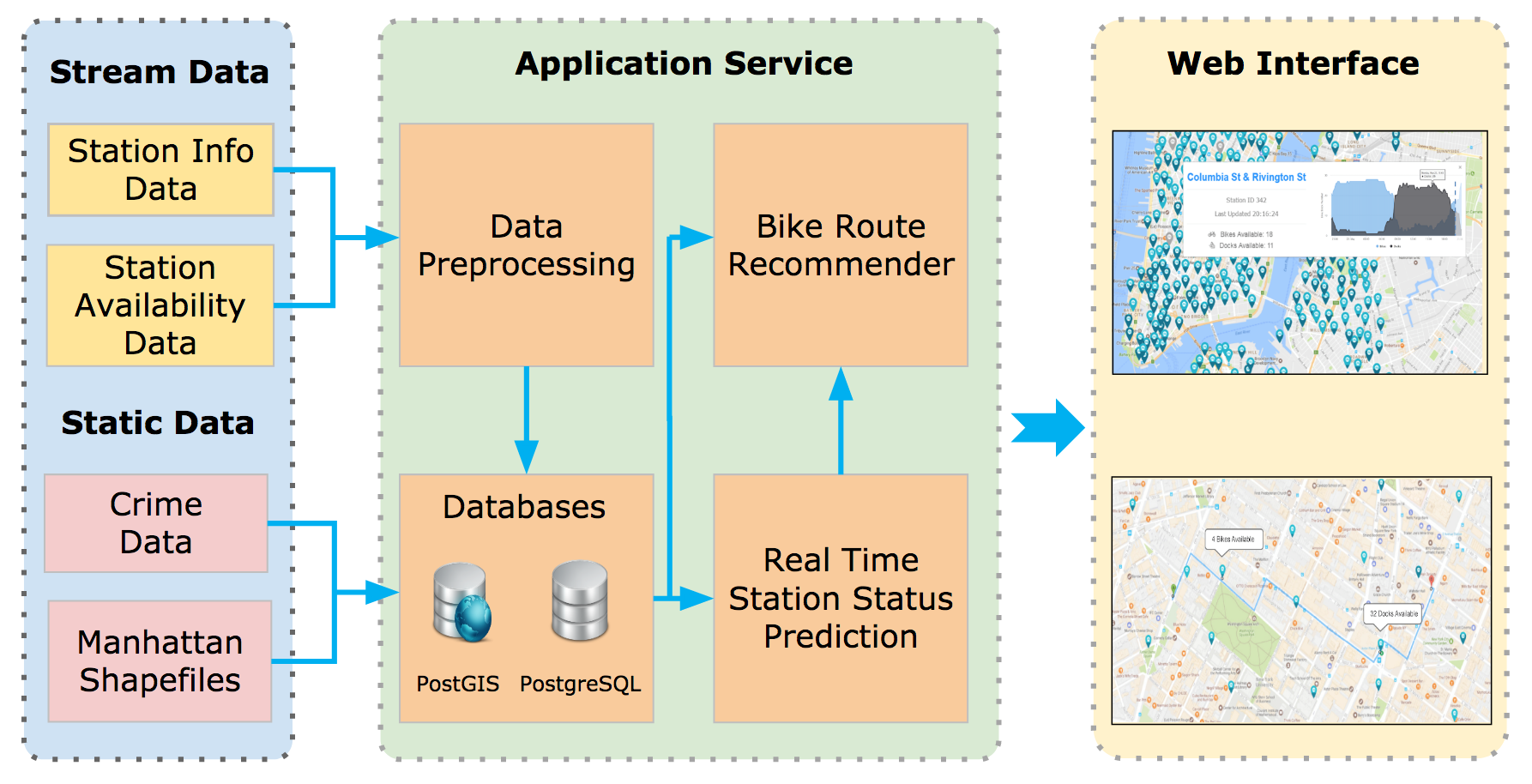}
\caption{System Architecture}
\end{figure}

\section{System Architecture}
In this section, the system architecture of SAFEBIKE is presented. At the high level, the system consists of three main components: data processing, application services, and the system interface, as shown in Figure 1. 

\subsection{Data Processing}
The major function of this component is to extract, pre-process, and store three types of data into appropriate databases, i.e., station information, station availability and crime  data. For station information and availability data, we utilized Citi Bike System API\cite{citi} to crawl real-time station information and availability. Each feedback is represented as a JSON file, so a data preprocessor was implemented for transferring each file into tuples and store them into the database. The detailed process is described in Section 3.1. For crime data, we collected spatial crime data from NYC Open Data\cite{nyc}. A data preprocessor is implemented to filter out compromised records.

\subsection{Application Services}
Two databases are maintained in SAFEBIKE: a PostgreSQL database is used for bike-sharing datasets, while a PostGIS database is used for crime data and shapefiles for Manhattan area. According to users' interests to available bikes and docks number at bike stations in the near future, we implemented a real-time availability prediction algorithm to forecast bikes, and docks demand at station level. Users can benefit from the forecast results to choose origin and destination stations.

Based on city roadway data from PostGIS database, along with route ranking factors such as predicted station availability data and crime records, we implemented an optimal route recommendation service according to the individual user interests.

\subsection{System Interface}
User interactions and operations such as sending requests and receiving results from the back-end server are performed by this module. The web application utilizes Python Flask to construct the framework. All communications to the back-end web server are implemented by Ajax technology. Google Maps APIs are leveraged to enable location-based services such as bike station map display, station availability visualization, and route recommendation.

\section{Features}
SAFEBIKE is capable of crawling station availability from Citi Bike System.  Based on station availability data, a real-time station availability prediction algorithm is proposed in SAFEBIKE. We develop an optimal route recommendation service, which takes user's preferences as inputs. The recommendation service generates the optimal bike route which takes the number of bikes available at origin station, docks available at the destination station, and the safety level into consideration.

\subsection{Real-time Station Status Prediction}
Station availability is a real-time information that reflects the change of bikes and docks of all stations in BSSs. Station availability consists of labels indicating whether a particular station is allowed for renting or returning, and the number of disabled bikes or docks. A availability crawler is designed to crawl the Citi Bike System API for availability data regularly. This crawler works constantly and fetches availability data every 10 minutes. The raw data is represented as JSON file which consists of availability of all stations in the Citi Bike System. For each station availability, it is parsed, and then stored into database.

An station availability predictor is implemented in our system to infer the number of available bikes and docks at each station for future time periods, for example, 10 minutes, 30 minutes or 1 hour later. Station availability at current time is taken as inputs to improve the quality of predicted results. We proposed a real-time availability prediction algorithm to compute the number of available bikes and docks for each station. For each station $s$, firstly, we transfer all its historical availability records into two types of availability vectors:
\begin{equation}
\begin{split}
  Bike_{d,t} = (B_{t},B_{t+1}, B_{t+2}, ..., B_{t+n})^T \\
  Dock_{d,t} = (D_{t},D_{t+1}, D_{t+2}, ..., D_{t+n})^T
  \end{split}
\end{equation}

where $B_i$ and $D_i$ represents the number of bikes and docks in station $s$ at time $i$, $i \in T$, $n \in \mathbb{N}$ is the number of future time periods we want to predict, $d \in Dates$ is the date a availability was collected, $Dates$ is the set of dates. In $Bike_{d,t}$ and $Dock_{d,t}$, the time interval between two availability records is set to 10 minutes. 

Next, for each station $s$ and each time $t$, we compute the average of all its $Bike$ and $Dock$ vectors:
\begin{equation}
\begin{split} 
  AvgBike_{w,t} = \frac{\sum _{d \in Dates,W(d)=w} Bike_{d,t}}{|Dates|} \\
  AvgDock_{w,t} = \frac{\sum _{d \in Dates,W(d)=w} Dock_{d,t}}{|Dates|} 
  \end{split}
\end{equation}

where we utilize a Boolean value $w$ to separately consider different station behaviors during weekdays and weekends, where $w=0$ indicates a weekday, $w=1$ indicates a weekend, W(d) is a function that $W(d)=0$ if $d$ is a weekday, and $W(d)=1$ if $d$ is a weekend.

Finally, given the bikes availability $B$ and docks availability $D$ of station $s$ at current time $t_c$ and corresponding current date $d_c$, the algorithm computes the predicted bikes and docks vectors as:
\begin{equation}
\begin{split}
PredBike_{d_c,t_c}  =  (PB_{t_c+1},PB_{t_c+2}, ...,PB_{t_c+n})^T \\
\text{where} \: PB_{t_c+i} = B_{t_c} + (AvgB_{W(d_c),t_c+i}-AvgB_{W(d_c),t_c}),  \\
AvgB_{W(d_c),t_c+i},AvgB_{W(d_c),t_c} \in AvgBike_{W(d_c),t_c}
\end{split}
\end{equation}
and
\begin{equation}
\begin{split}
PredDock_{d_c,t_c}  =  (PD_{t_c+1},PD_{t_c+2}, ...,PD_{t_c+n})^T \\
\text{where} \: PD_{t_c+i} = D_{t_c} + (AvgD_{W(d_c),t_c+i}-AvgD_{W(d_c),t_c}),\\
AvgD_{W(d_c),t_c+i},AvgD_{W(d_c),t_c} \in AvgDock_{W(d_c),t_c}
\end{split}
\end{equation}

where $PredBike$ and $PredDock$ are the predict vectors of bikes and docks in future time periods, respectively. The forecast number of bikes/docks at $i^{th}$ time interval after current time $t_c$ is represented as $PB_{t_c+i}$/$PD_{t_c+i}$, respectively, $0<i \leq n$. To be specific, $PB_{t_c+i}$/$PD_{t_c+i}$ is computed as the sum of current bikes/docks availability $B_{t_c}$/$D_{t_c}$ and average change of bikes/docks availability from current time $t_c$ to future time $t_c+i$.

\subsection{Routing Schemes}
Users provide expected departure time and the locations of origin and destination as input parameters for the route recommendation service. The system suggests an optimal route between two locations that considers three factors: length, safety, and the availability of bikes and docks at origin and destination stations. Users are allowed to customize the weights (factor distribution) of these factors to adjust the relative influence of each. Inspired by the existing shortest routing algorithms, we develop our routing algorithms based on the Dijkstra's algorithms.

Different from traditional routes, since users aim to travel from origin to destination using bike-sharing system, our routes includes two bike stations that users must go through: origin station and destination station. Therefore, in our system, a route is a combination of three components: a walking sub-route from start point to origin station, a biking sub-route from origin station to destination station and another walking sub-route from destination station to destination point. We perform our routing algorithm separately on these three sub-routes, and then combine them together into a entire route. To constrain the total walking distance, we only consider stations within distance-k buffer area of origin and destination.

The problem of finding the optimal route is projected onto an optimization problem that minimizes the cost of the route\cite{Fu:2014:TSR:2666310.2666368}. The edge cost weighting is determined from two perspectives: distance factor and crime factor. For simplicity, we consider users have same preferences on walking route and biking route. For the cost of entire route, as we also consider the availability of bikes and docks in the origin and destination stations, the cost is determined by the combination of above two factors and station availability factor. By adjusting the weights of these three factors appropriately, we propose three routing schemes: shortest route, safest route and optimal route.

\textbf{A. Shortest Route Scheme}
Distance is the basic factor for most routing systems. By considering this factor, the edge is weighted only by the length of the trajectory. The objective function is:
\begin{equation}
  \min_{O \xrightarrow{E} D} \big\{ \sum_{e \in E} length(e) \big\}
\end{equation}

where $O$ and $D$ are the origin and destination predefined by the user. The function identifies a set of trajectories with the shortest total lengths.

\textbf{B. Safest Route Scheme}
The ability to incorporate crime data is an important aspect of SAFEBIKE. Since an assumption has been made that users of our application will treat safety as a relatively important factor while they are traveling, the system takes this factor into consideration and recommends the safest route. This factor weights the trajectories by the number of crime reports within its distance-d buffer area. The objective function is: 
\begin{equation}
  \min_{O \xrightarrow{E} D} \big\{ \sum_{e \in E} crime(e) \big\}
\end{equation}

where $crime(e)$ represents the accumulated number of crime incidents within distance-d buffer area of road segment $e$. The function identifies a set of trajectories with the fewest accumulated crime incidents.

\textbf{C. Optimal Route Scheme}
This scheme combines the above two edge weighting schemes and stations availability factor by specifying three weighting parameters: $\alpha, \beta$ and $\gamma$. The objective function for this scheme is: 
\begin{equation}
  \min_{O \xrightarrow{E} D} \big\{ \sum_{e \in E} \big( \alpha \cdot nlength(e) + \beta \cdot ncrime(e) \big) + \gamma \cdot (1-nAVL(E)) \big \}
\end{equation}

where $nlength(e)$ is the normalized edge length, $ncrime(e)$ is the normalized crime number, and $nAVL(E)$ is the normalized  bikes and docks availability for route $E$. That is to say: 
\begin{align*}
    nlength(e) &= length(e)/ max_{e^\prime \in C}\{length(E^\prime)\} \text{ and} \\
    ncrime(e) &= crime(e)/ max_{e^\prime \in C}\{crime(E^\prime)\} \text{ and} \\
    nAVL(E) &= AVL(E)/ max_{e^\prime \in C}\{AVL(E^\prime)\}
\end{align*}

where $E^\prime$ is one of the route in candidate route set $C$, $AVL(E) =PB_{t_{out},i} \cdot PD_{t_{in},j}$ represents the station availability availability of route $E$, $PB_{t_out,i}$ is the predicted number of bikes in route $E$'s origin station $i$ at predicted check-out time $t_{out}$, and $PD_{t_in,j}$ is the predicted number of docks in route $E$'s destination station $j$ at predicted check-in time $t_{in}$. The factor distribution [$\alpha, \beta, \gamma$] is a three dimensional vector which is located on the plane $\alpha+ \beta+ \gamma =1$, where $\alpha, \beta, \gamma \in$ [0,1], and the previous two schemes are special cases when $\alpha=1$ and  $\beta=1$, respectively.

\section{Demonstration Scenario}
In this section, we compare routing performance under different factor distributions, and a real case study is introduced to show the effectiveness of our system.
\subsection{Routing Performance Comparison}
In Figure 2, the red, green and blue lines are the shortest route, safest route and the optimal route, respectively. The blue markers are locations for bike stations, the dark blue and light blue colors represent the ratio between bikes and docks. The vector [$\alpha, \beta, \gamma$] is the factor distribution. By default, the vector is set to [0.3,0.3,0.4] and the corresponding result is shown in Figure 2(b).

Although the red route is the shortest, it does not take users through safe areas. The green route goes a longer way to the destination, but it has the minimum number of crime incidents along the path. Both these two paths do not consider bikes and docks availability in stations, so users may have no bike to borrow or no dock to return the bike from the recommended stations. The optimal path is the result of trade-offs between length, safety and bikes/docks availability. By changing the factor distribution of the optimal route, it approaches the shortest route and choose a origin station with more available bikes when length factor $\alpha$ is large (Figure 2(a)) and the safest path when $\beta$ is large (Figure 2(d)). When $\gamma$ is large (Figure 2(c)), optimal path choose origin station with many bikes and destination station with many docks available, but the user needs to walk a longer distance before picking up a bike and after dropping off the bike.
\begin{figure}[!htb]
\centering
\subfigure[$\alpha, \beta, \gamma$ = {[0.8,0.1,0.1]} ]{\centering \includegraphics[width=0.45\linewidth,height=0.45\linewidth]{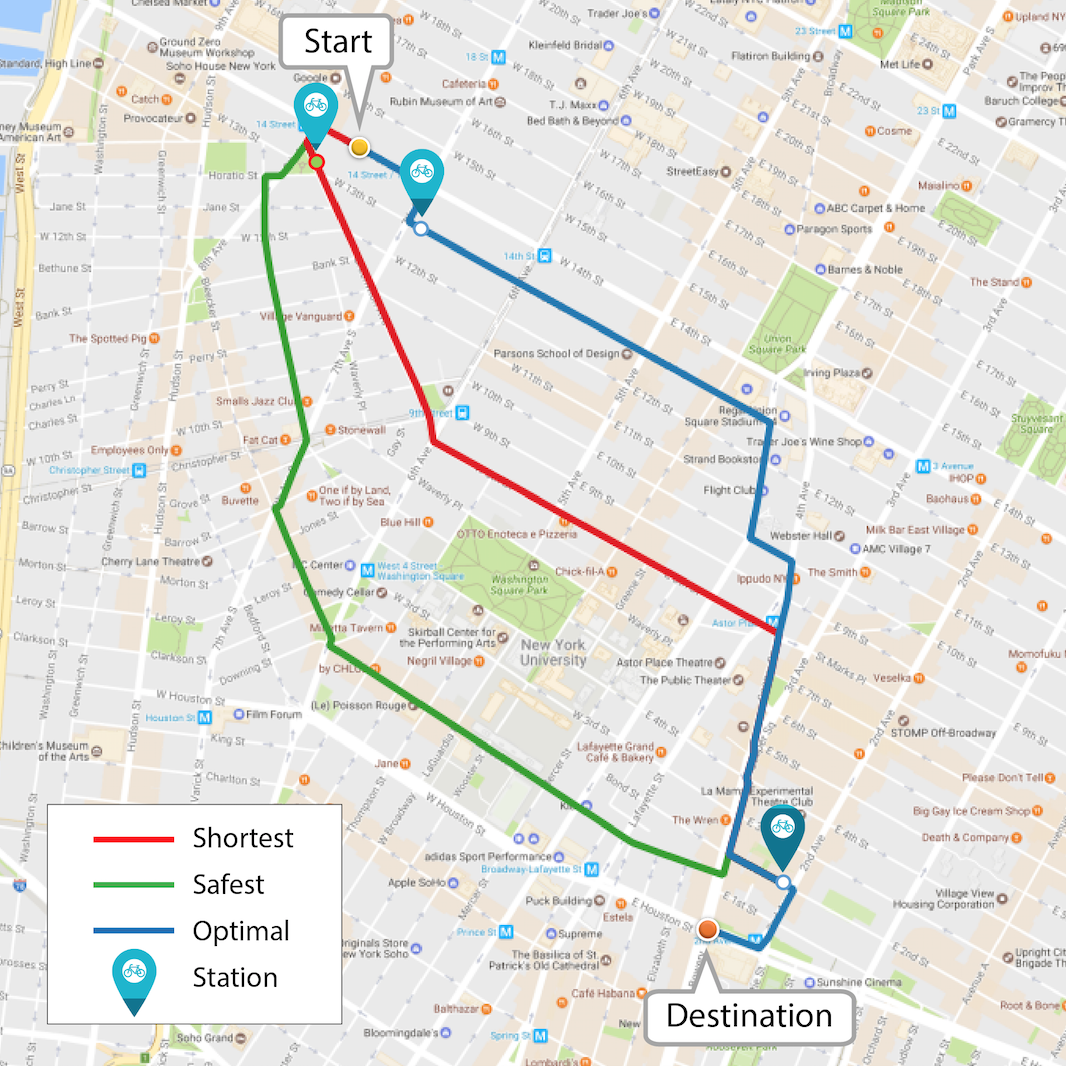}}
\subfigure[$\alpha, \beta, \gamma$ = {[0.3,0.3,0.4]} ]{\centering \includegraphics[width=0.45\linewidth,height=0.45\linewidth]{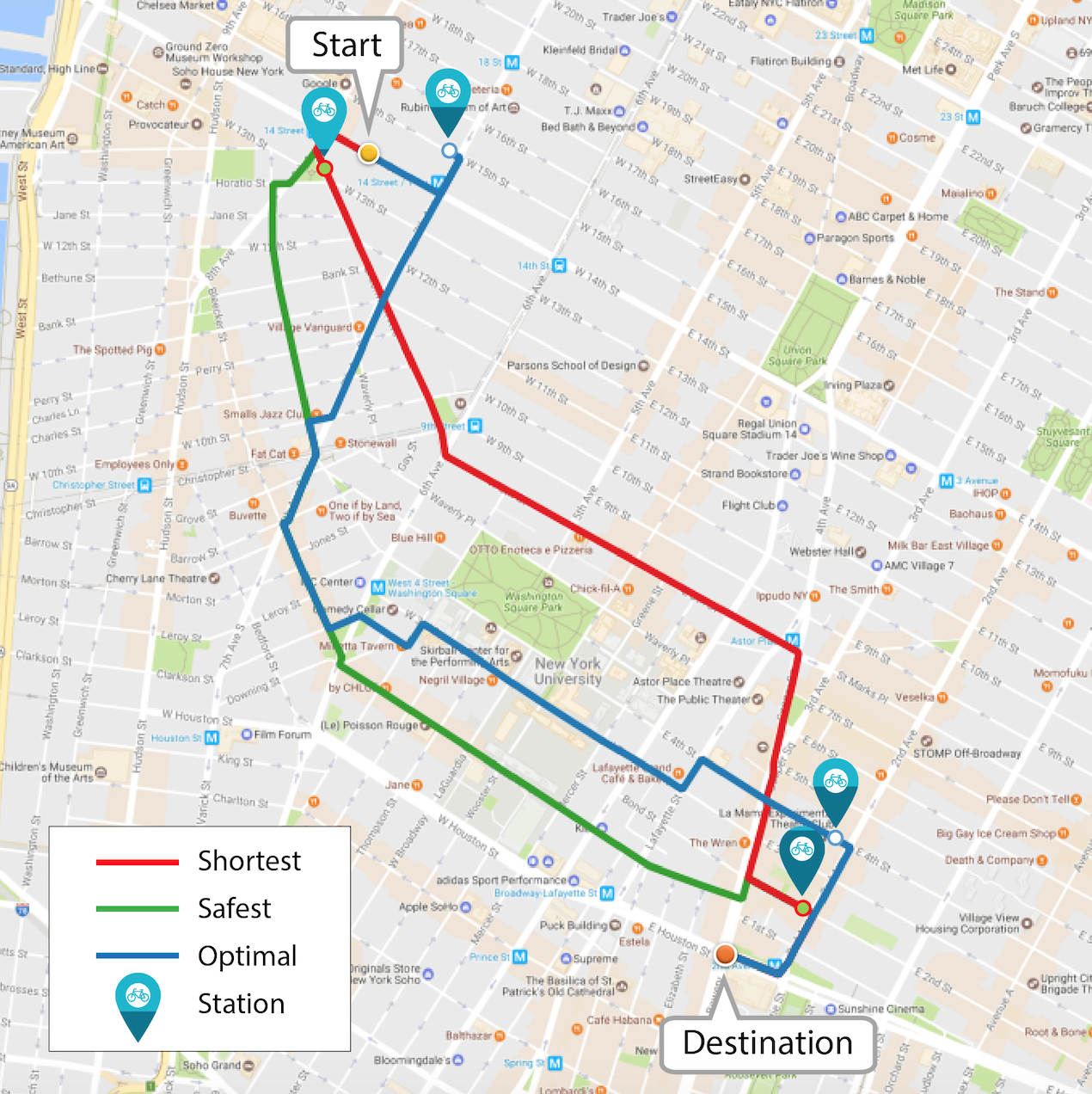}}
\subfigure[$\alpha, \beta, \gamma$ = {[0.1,0.1,0.8]} ]{\centering \includegraphics[width=0.45\linewidth,height=0.45\linewidth]{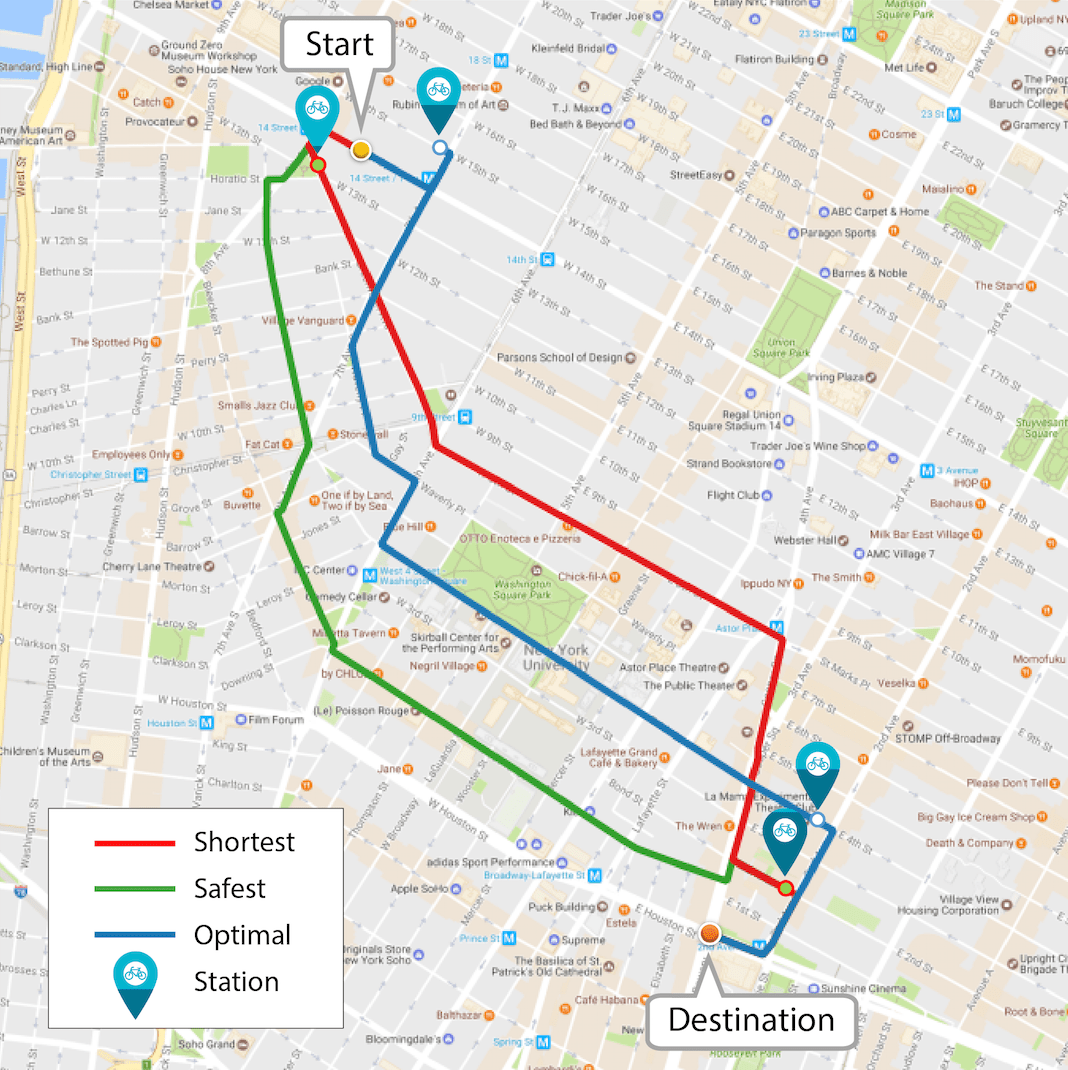}}
\subfigure[$\alpha, \beta, \gamma$ = {[0.1,0.8,0.1]} ]{\centering \includegraphics[width=0.45\linewidth,height=0.45\linewidth]{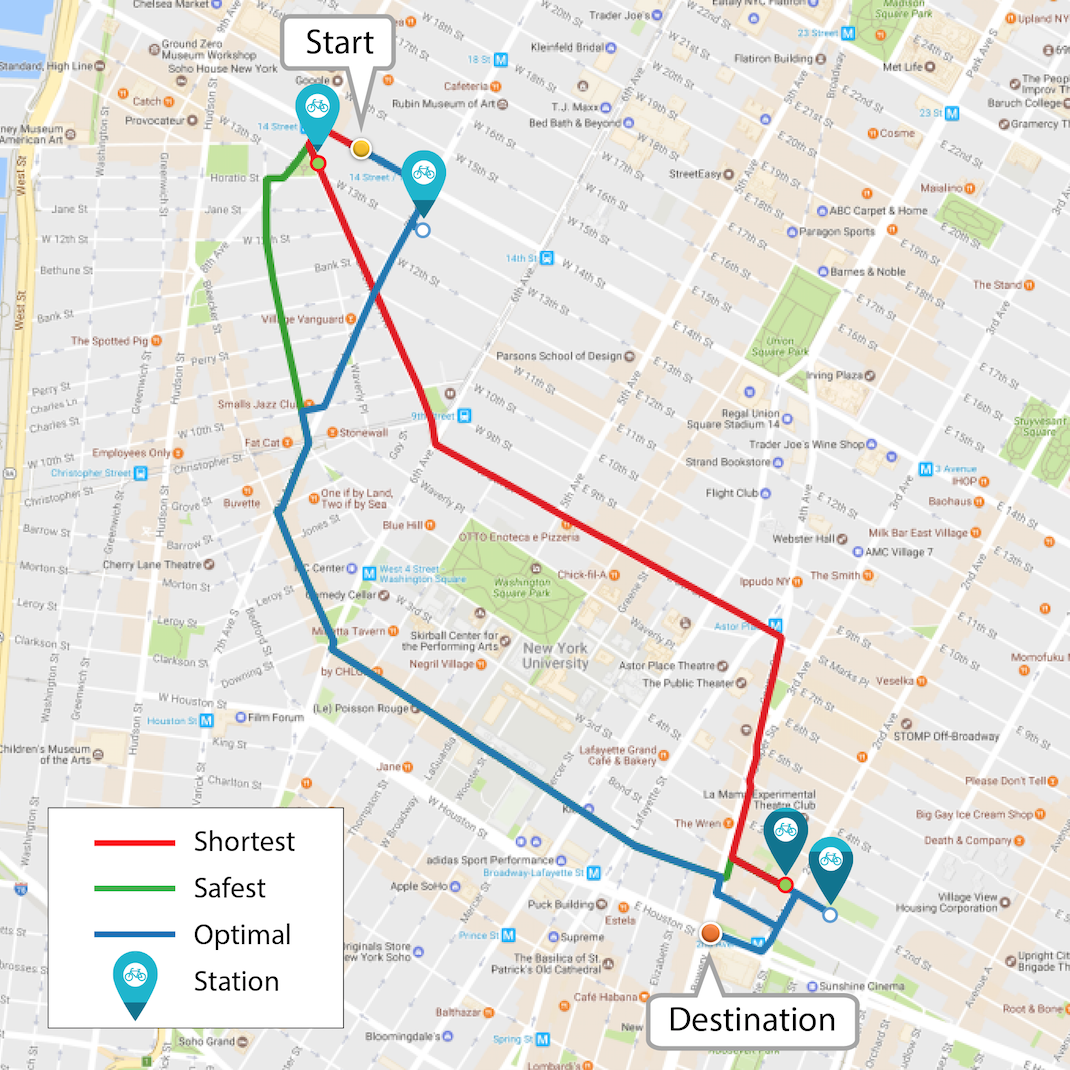}}
\caption{Routing Results for Different Factor Distributions}
\end{figure}
\subsection{System Interface}
SAFEBIKE user interface is designed and demonstrated using NYC Citi Bike database. The main user interface is shown in Figure 3, and the light blue and dark blue colors indicate the different availability of stations across the City. For each station, an infobox presents the current availability of this station, while a chart shows historical availability in the last 24 hours with predicted availability in the next hour.
\begin{figure}[!htb]
\centering
\includegraphics[width=1\linewidth,height=4cm]{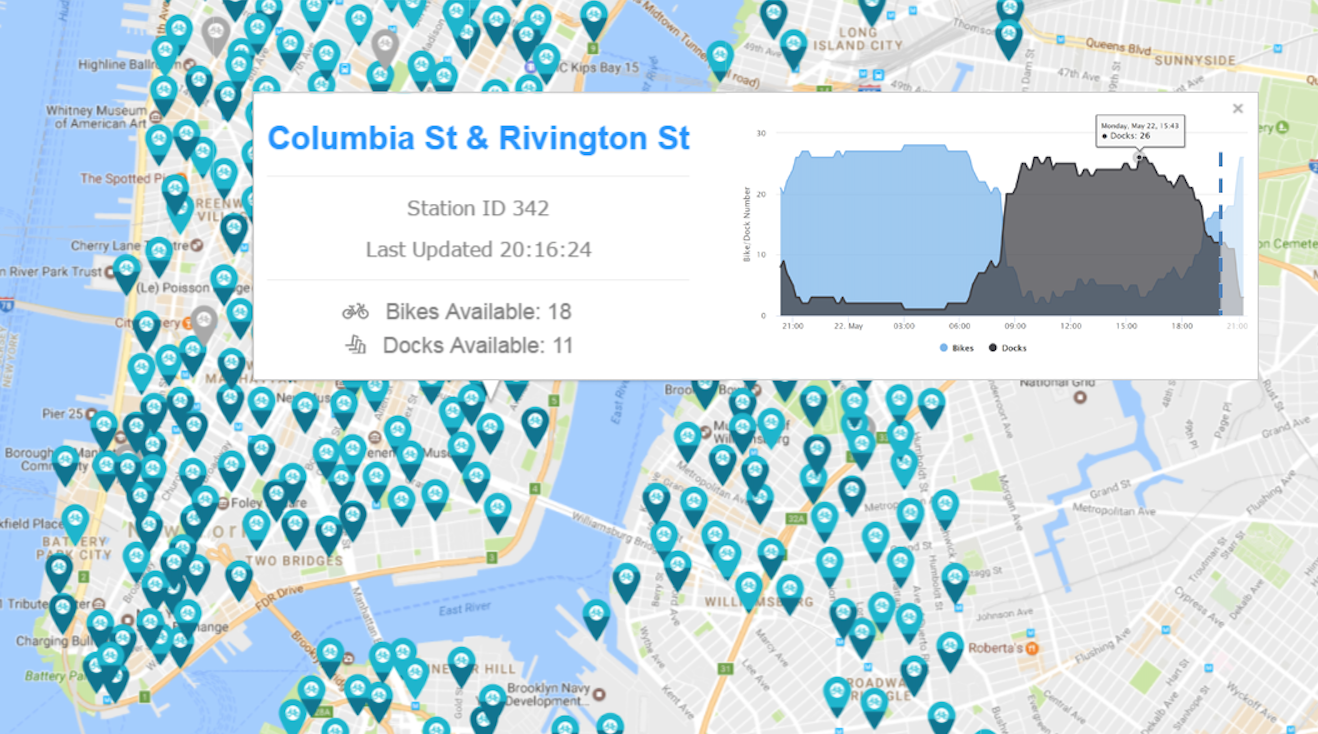}
\caption{System Interface for May 22, 2017}
\end{figure} 
\subsection{Case Study}
\begin{figure}[!htb]
\centering
\subfigure[Shortest Route]{\centering \includegraphics[width=0.95\linewidth,height=2.3cm]{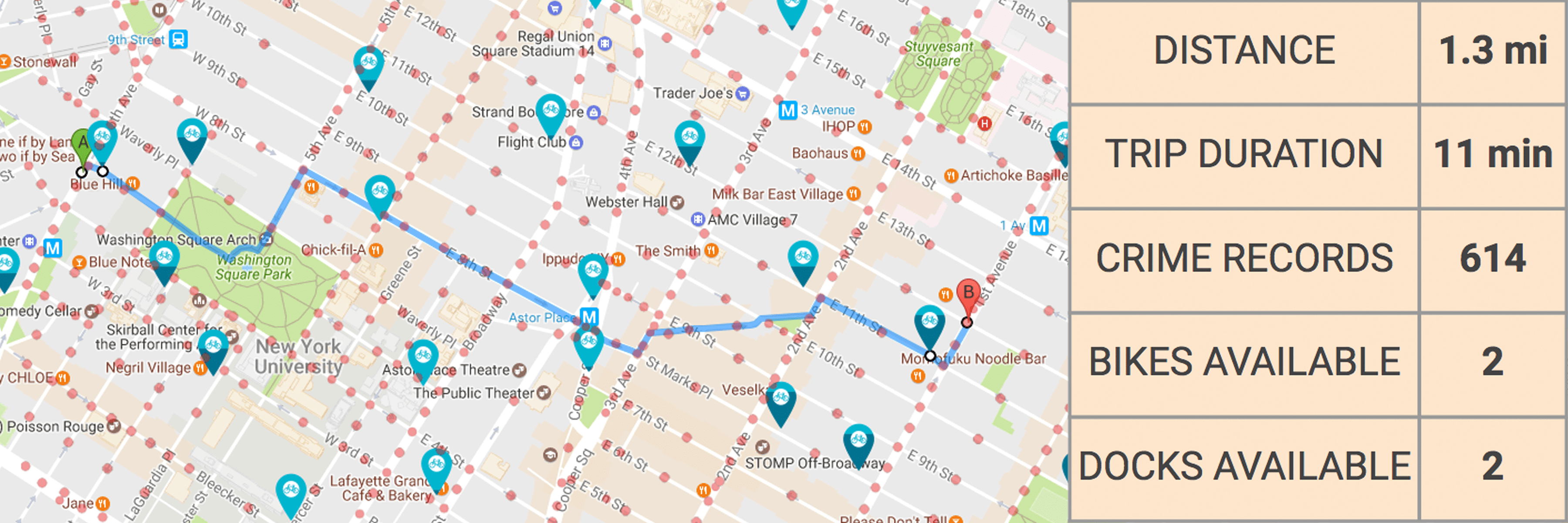}}
\subfigure[Safest Route]{\centering \includegraphics[width=0.95\linewidth,height=2.3cm]{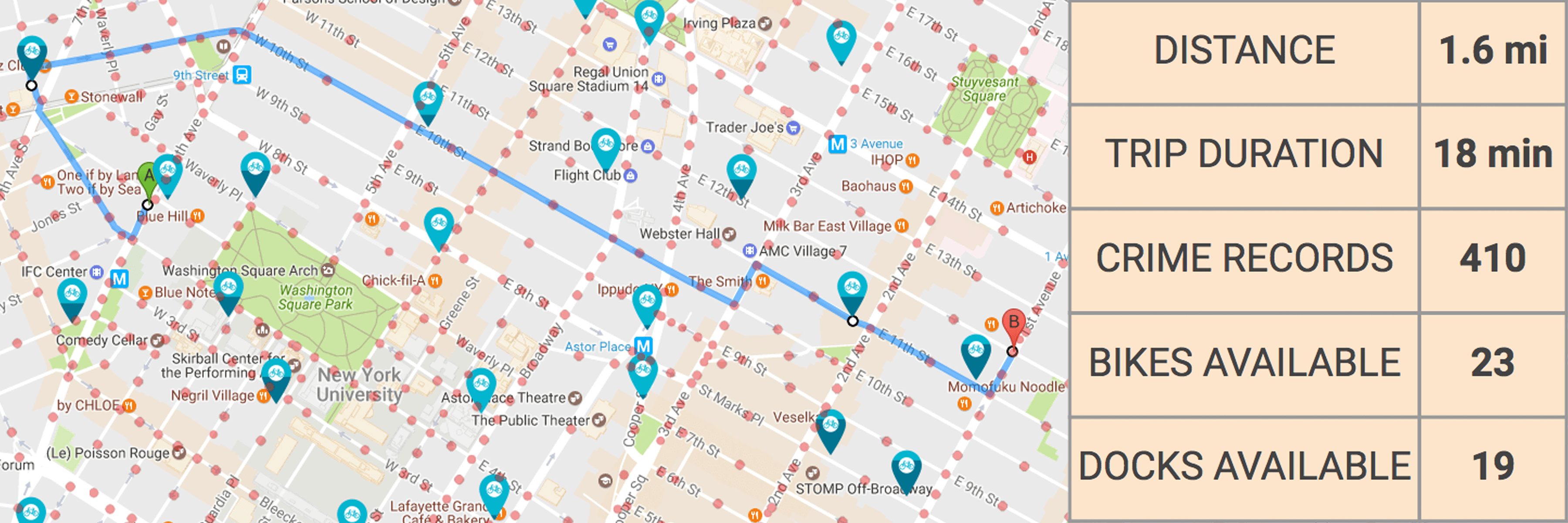}}
\subfigure[Optimal Route]{\centering \includegraphics[width=0.95\linewidth,height=2.3cm]{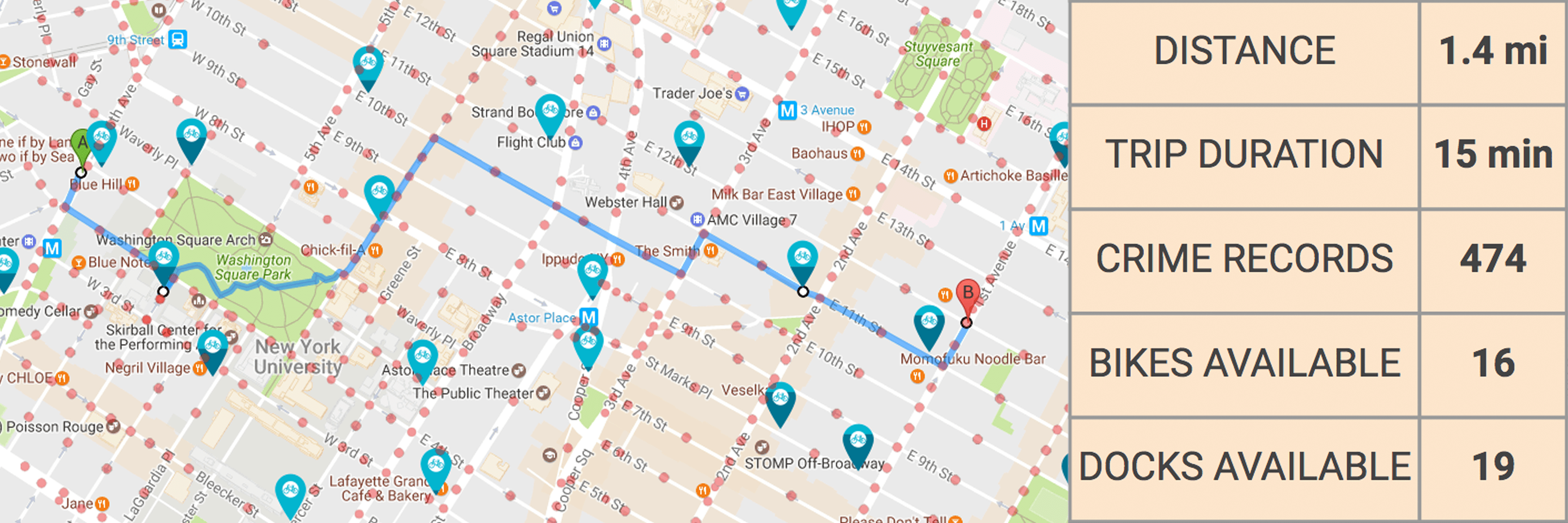}}
\caption{Three Cases of Recommended Routes}
\end{figure}
Figure 4 presents three recommended routes from the intersection of \emph{6th Ave} and \emph{Washington PI} to the intersection of \emph{1st Ave} and \emph{E 12th St} at 2:27 PM on a Sunday. Figure 4(a) shows the shortest route, which takes only 11 minutes with no safety and bikes/docks availability concerns. Although its trip duration is the shortest, there are more crime incidents along the route. Also, the user may have a problem with checking out bike at the recommended origin station with only 2 bikes left, and the same for the destination station. The safest route in Figure 4(b) minimizes the number of crime incidents along the route but its duration (18 minutes) is  much more than that of the shortest route. As shown in Figure 4(c), the optimal route is the best among these three routes, which suggests the user to pick up a bike at the station that have 16 bikes available and return the bike to the station that has 19 docks available. It overlaps with part of the safest route, indicates the optimal route takes safeness into consideration.The case study shows that our application is very useful and practical in assisting users with bike trip planning.

\section{Conclusion}
SAFEBIKE is designed to meet the needs of end users who need to find the most convenient bike route between desired locations while avoiding areas with high crime rate. SAFEBIKE implements real-time station availability prediction algorithm to forecast future bikes and docks availability to provide a prior knowledge of station availability to users before they make a decision. The optimal router feature facilitates users to plan for safe and convenient bike trips. SAFEBIKE is envisioned to eventually extend to multiple cities with bike-sharing systems and enhance the overall life quality of bike-sharing users.


\begin{thebibliography}{00}


\ifx \showCODEN    \undefined \def \showCODEN     #1{\unskip}     \fi
\ifx \showDOI      \undefined \def \showDOI       #1{#1}\fi
\ifx \showISBNx    \undefined \def \showISBNx     #1{\unskip}     \fi
\ifx \showISBNxiii \undefined \def \showISBNxiii  #1{\unskip}     \fi
\ifx \showISSN     \undefined \def \showISSN      #1{\unskip}     \fi
\ifx \showLCCN     \undefined \def \showLCCN      #1{\unskip}     \fi
\ifx \shownote     \undefined \def \shownote      #1{#1}          \fi
\ifx \showarticletitle \undefined \def \showarticletitle #1{#1}   \fi
\ifx \showURL      \undefined \def \showURL       {\relax}        \fi
\providecommand\bibfield[2]{#2}
\providecommand\bibinfo[2]{#2}
\providecommand\natexlab[1]{#1}
\providecommand\showeprint[2][]{arXiv:#2}

\bibitem[\protect\citeauthoryear{Borgnat, Abry, Flandrin, Robardet, Rouquier,
  and Fleury}{Borgnat et~al\mbox{.}}{2011}]%
        {borgnat2011shared}
\bibfield{author}{\bibinfo{person}{Pierre Borgnat}, \bibinfo{person}{Patrice
  Abry}, \bibinfo{person}{Patrick Flandrin}, \bibinfo{person}{C{\'e}line
  Robardet}, \bibinfo{person}{Jean-Baptiste Rouquier}, {and}
  \bibinfo{person}{Eric Fleury}.} \bibinfo{year}{2011}\natexlab{}.
\newblock \showarticletitle{Shared bicycles in a city: A signal processing and
  data analysis perspective}.
\newblock \bibinfo{journal}{{\em Advances in Complex Systems\/}}
  \bibinfo{volume}{14}, \bibinfo{number}{03} (\bibinfo{year}{2011}),
  \bibinfo{pages}{415--438}.
\newblock


\bibitem[\protect\citeauthoryear{Data}{Data}{2017}]%
        {nyc}
\bibfield{author}{\bibinfo{person}{NYC~Open Data}.}
  \bibinfo{year}{2017}\natexlab{}.
\newblock \bibinfo{title}{{Open Data for All New Yorkers}}.
\newblock   (\bibinfo{year}{2017}).
\newblock
\showURL{%
\url{https://opendata.cityofnewyork.us/}}


\bibitem[\protect\citeauthoryear{Fishman and Schepers}{Fishman and
  Schepers}{2016}]%
        {fishman2016global}
\bibfield{author}{\bibinfo{person}{Elliot Fishman} {and} \bibinfo{person}{Paul
  Schepers}.} \bibinfo{year}{2016}\natexlab{}.
\newblock \showarticletitle{Global bike share: what the data tells us about
  road safety}.
\newblock \bibinfo{journal}{{\em Journal of safety research\/}}
  \bibinfo{volume}{56} (\bibinfo{year}{2016}), \bibinfo{pages}{41--45}.
\newblock


\bibitem[\protect\citeauthoryear{Fu, Lu, and Lu}{Fu et~al\mbox{.}}{2014}]%
        {Fu:2014:TSR:2666310.2666368}
\bibfield{author}{\bibinfo{person}{Kaiqun Fu}, \bibinfo{person}{Yen-Cheng Lu},
  {and} \bibinfo{person}{Chang-Tien Lu}.} \bibinfo{year}{2014}\natexlab{}.
\newblock \showarticletitle{TREADS: A Safe Route Recommender Using Social Media
  Mining and Text Summarization}. In \bibinfo{booktitle}{{\em Proceedings of
  the 22Nd ACM SIGSPATIAL International Conference on Advances in Geographic
  Information Systems}} {\em (\bibinfo{series}{SIGSPATIAL '14})}.
  \bibinfo{publisher}{ACM}, \bibinfo{address}{New York, NY, USA},
  \bibinfo{pages}{557--560}.
\newblock
\showISBNx{978-1-4503-3131-9}
\showDOI{%
\url{https://doi.org/10.1145/2666310.2666368}}


\bibitem[\protect\citeauthoryear{System}{System}{2017}]%
        {citi}
\bibfield{author}{\bibinfo{person}{CitiBike System}.}
  \bibinfo{year}{2017}\natexlab{}.
\newblock \bibinfo{title}{{General Bikeshare Feed Specification}}.
\newblock   (\bibinfo{year}{2017}).
\newblock
\showURL{%
\url{https://github.com/NABSA/gbfs/blob/master/gbfs.md}}


\bibitem[\protect\citeauthoryear{Wikipedia}{Wikipedia}{2017}]%
        {wiki}
\bibfield{author}{\bibinfo{person}{Wikipedia}.}
  \bibinfo{year}{2017}\natexlab{}.
\newblock \bibinfo{title}{{List of Bicycle-Sharing Systems}}.
\newblock   (\bibinfo{year}{2017}).
\newblock
\showURL{%
\url{https://en.wikipedia.org/wiki/List_of_bicycle-sharing_systems}}


\end{thebibliography}

\end{document}